\documentclass[prl,twocolumn,showpacs,groupedaddress]{revtex4}

\usepackage{dcolumn}
\usepackage{graphicx}
\usepackage{amsmath}
\usepackage{bm}

\usepackage[usenames]{color}

\begin{document}


\title{Spontaneous Conversion from Virtual to Real Photons \\ in the Ultrastrong Coupling Regime}


\author{R. Stassi$^1$, A. Ridolfo$^2$, O. Di Stefano$^1$, M. J. Hartmann$^2$, and S. Savasta$^1$}
\affiliation{$^1$Dipartimento di Fisica e Scienze della Terra, Universit\`{a} di Messina, Viale F. Stagno d'Alcontres 31, I-98166 Messina, Italy\\
$^2$Physik Department, Technische Universit\"{a}t M\"{u}nchen, 85748 Garching, Germany} 
\begin{abstract}
{
We show that
a spontaneous release of virtual photon pairs can occur in a quantum optical system in the ultrastrong coupling regime. In this regime, which is attracting interest both in semiconductor and superconducting systems, the light-matter coupling rate $\Omega_{\rm R}$ becomes comparable to the bare resonance frequency of photons $\omega_0$.
In contrast to the dynamical Casimir effect and other pair creation mechanisms, this phenomenon does not require external forces or time dependent parameters in the Hamiltonian.}
\end{abstract}


\maketitle
One of the most surprising predictions of modern quantum theory
is that the vacuum of space is not empty but  filled with a {\em sea} of virtual particles.
These short-lived fluctuations are the origin of some of the
most important physical processes in the universe.
A quite direct evidence of the existence of such virtual particles is provided by the dynamical Casimir effect (DCE).
It predicts that rapid modulations of the boundary conditions of a quantum field induce vacuum amplification effects that result in the creation of real particles out of vacuum fluctuations.
The DCE \cite{moore} has been recently experimentally realized  in superconducting circuits \cite{Wilson, Hakonen}. Other proposed vacuum amplification mechanisms  \cite{Nation}, as the Schwinger process \cite{Schwinger}
and the Hawking radiation \cite{Hawking75}, require the presence of huge external fields  or, as the Unruh effect, the presence of a rapidly nonuniformly accelerating observer \cite{Unruh}, and  still await observation.
In this Letter we consider a three level emitter where the transition between the two upper levels couples ultrastrongly to a cavity mode and show that the spontaneous relaxation of the emitter from its intermediate to its ground state is accompanied by the creation of photons in the cavity mode (see Fig.\ 1).

The Hamiltonian of a realistic atom-cavity system contains so-called counter-rotating terms allowing the simultaneous creation or annihilation of an excitation in both atom and cavity mode. These terms can be safely neglected for small coupling rates $\Omega_{\rm R}$ in the so called rotating-wave approximation (RWA). However, when $\Omega_{\rm R}$
becomes comparable to the cavity resonance frequency of the emitter  or the resonance frequency of the cavity mode, the counter-rotating terms are expected to manifest, giving rise to exciting effects in cavity-QED \cite{DeLiberato1, Niemczyk, Casanova, RidolfoPRL2013}. 
This ultrastrong-coupling (USC) regime is difficult to reach in quantum-optical
cavity-QED, but was recently realized in a variety of solid-state quantum systems \cite{guenter09,Niemczyk,Schwartz,Hoffman, Scalari}.
Such regime is challenging from a theoretical point of view as the total number of
excitations in the cavity-emitter system is not preserved, even though its parity is \cite{Casanova}.
It has been shown that, in the USC regime, the quantum optical master equation fails to provide the correct description of the system's interaction with reservoirs \cite{Blais}.
Moreover quantum optical normal order correlation functions fail to describe photodetection experiments for such systems \cite{Savasta96,DeLiberato2}. 
Specifically, for a single mode resonator, the photon rate that can be detected by a photoabsorber is no longer proportional
to $\langle a^\dag(t) a(t) \rangle$, where $a$ and $a^{\dagger}$ are the photon destruction and creation operators of the cavity mode, but to $\langle X^-(t) X^+(t) \rangle$, where  $X^+(t)$ is the positive frequency component of the quadrature operator $X(t) = a(t) + a^\dag(t)$ \cite{RidolfoPrl2012}.
A puzzling property of these systems is that their ground state is a squeezed
vacuum containing correlated pairs of cavity photons \cite{Ashhab}.
\begin{figure}[!ht]  
\includegraphics[height= 26 mm]{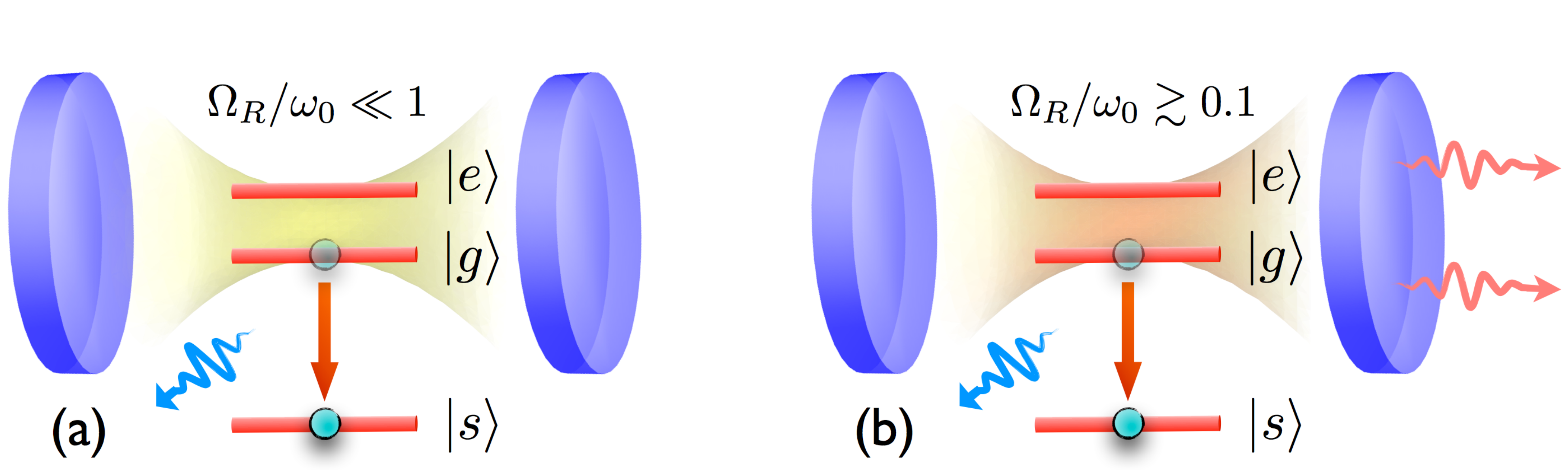}  
\caption{(color online)  Sketch of the system and of  the process under consideration. Two-levels of a single cascade three-level system  are  coupled
to a single cavity mode. Sketch of the spontaneous decay process $| g \rangle \to |s \rangle$ in the case of weak and strong coupling ($\Omega_{\rm R}/\omega_0 \ll 1$) (a), and in the USC regime ($\Omega_{\rm R}/\omega_0 \gtrsim 0.1$) (b). In the latter case the spontaneous decay can produce the emission of additional pairs of cavity photons.
} 
\label{fig:scheme}
\end{figure}
The photon pairs in this ground state $| \tilde{0}\rangle$ are, however, virtual and cannot be detected \cite{Ciuti}, as $\langle \tilde{0} |X^-(t) X^+(t) | \tilde{0} \rangle = 0$ \cite{RidolfoPrl2012}. Otherwise, an observation
of a stream of photons from such a system in its ground state would give rise to {\em perpetuum mobile} behaviors. Nevertheless, we show that if the two-level transition ultrastrongly coupled with the resonator is built using the excited states of a three-level
artificial atom (see Fig.\ 1), such virtual photons can spontaneously convert into actually measurable photons that are uncoupled from the atom.
Such unconventional spontaneous emission can be traced to the  structure of the electrodynamic vacuum in the USC regime. 

A mechanism for the generation of quantum vacuum radiation, based on the presence of counter-rotating terms in the light-matter interaction Hamiltonian, has already been proposed \cite{Bastard, DeLiberato1, DeLiberato2}. 
All these vacuum amplification proposals, as the DCE, require a fast  time-modulation or the sudden switch on or off of the vacuum Rabi frequency.
The most promising candidates for an experimental
realization of the proposed spontaneous conversion effect (SCE) are superconducting quantum circuits \cite{Norireview,Niemczyk} and intersubband quantum well polaritons \cite{guenter09}. In particular experimental realizations of  circuit-QED
systems operating in the USC regime were achieved by using single phase-biased flux artificial atoms \cite{Niemczyk,Forn}.
By adjusting the externally applied reduced magnetic flux, these  artificial atoms can acquire the quantized  level structure (see Fig.\ 1) as well as the transition matrix elements  required for the observation of the SCE \cite{suppl}. 
Suitable systems that reach the USC are intersubband transitions in undoped quantum wells embedded in a microcavity, although
such a regime is achieved with the contribution of a very large number of effective quantum emitters.
In addition, such a system can display the level structure required for the observation of the SCE \cite{guenter09, suppl}.

Let us consider a three-level cascade quantum system with the ground state level labeled as $|{\rm s} \rangle$ and the first and second excited states
respectively  $|{\rm g} \rangle$ and $|{\rm e} \rangle$. We treat $|{\rm g} \rangle$ and $|{\rm e} \rangle$ as the two-level system resonantly coupled to the resonator (see Fig.\ 1). Interesting theoretical studies of quantum dynamics in these cavity-QED systems with \cite{RidolfoPRL2011,Stassi2012} and without \cite{backaction}  the RWA recently appeared.
In the absence of losses, the emitter-resonator  system is described
by the  Hamiltonian
\begin{equation}\label{eq:model}
    H = \omega_{\rm 0} a^{\dagger}a + \sum_{\alpha= {\rm s, g,e}} \omega_\alpha \sigma_{\alpha \alpha}  + \Omega_{\rm R} ( a + a^{\dag})(\sigma_{\rm eg}+ 
    \sigma_{\rm ge})\, ,
\end{equation}
where  $\omega_{0}$ is the frequency of the cavity mode, $\omega_{\alpha}$ ($\alpha={\rm s,g,e}$) are the bare frequencies of the atom-like relevant states, and $\sigma_{\alpha \beta} = |\alpha \rangle\langle \beta|$ describes the transition operators (projection operators if $\alpha= \beta$) involving the levels of the quantum emitter. It is useful to label with $| j \rangle$ ($j$ integer) the eigenstates of $H$, and with $\Omega_j$ (increasing with $j$)  the corresponding eigenenergies. The Hamiltonian $H$ describes a cascade three-level system with only one transition $|{\rm g}\rangle \leftrightarrow |{\rm e}\rangle$ resonantly coupled with a single mode of an optical resonator. Hence, when the system is in the state $|{\rm s}\rangle$, it does not interact with the resonator. The Hamiltonian can be split up as $H = H_{\rm Rabi} + H_{\rm s}$, where $H_{\rm Rabi}$ is the well known Rabi Hamiltonian and $H_{\rm s} = \omega_{\rm s}\sigma_{\rm s s}$.
As a consequence the total Hamiltonian is block-diagonal and its eigenstates can be separated into (i) a non-interacting sector $|{\rm s}, n \rangle$, with energy $\omega_{\rm s}+ n \omega_0$, where $n$ labels the cavity photon number; and into (ii) dressed atom-cavity states $| \tilde{j} \rangle$, resulting from the diagonalization of the Rabi Hamiltonian $H_{\rm Rabi}$. All subsequent calculations are performed at zero detuning: $\omega_0 = \omega_{\rm eg}$, being $\omega_{\alpha \beta} = \omega_{\alpha} - \omega_{\beta}$.
\begin{figure}[!ht]  
\includegraphics[height= 36 mm]{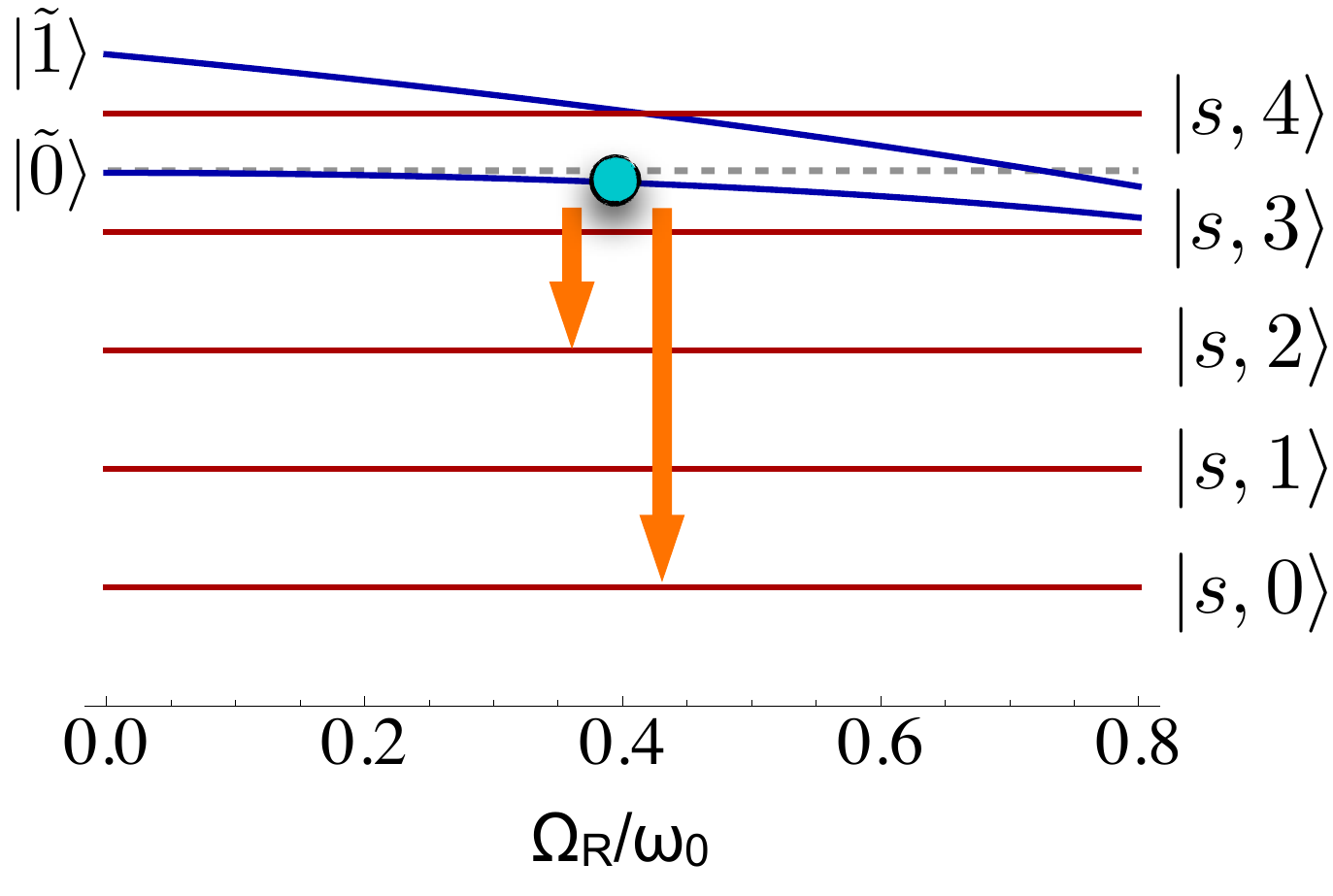}  
\caption{(color online)
Lowest energy levels of the total quantum system (eigenvalues of $H$) as a function of the coupling $\Omega_{\rm R}/\omega_0$ and a sketch of the allowed spontaneous transitions in USC. The  bending lines as a function of
$\Omega_{\rm R}/\omega_0$ describes the dressed energy levels $\omega_{\tilde j}$.}
\label{fig:2}
\end{figure}
Figure\ 2 displays the lowest energy levels (resulting from the numerical diagonalization of $H$) as a function of the normalized coupling $\Omega_{\rm R}/ \omega_0$. The equally energy-spaced flat lines (red) correspond to eigenstates of the non-interacting sectors $|{\rm s}, n \rangle$. The energy levels (blue) that split and then bend as a function of the $\Omega_{\rm R}/ \omega_0$  correspond to the eigenvectors of the interacting sectors $| \tilde{j} \rangle$. The lowest state of this sector can be expanded in terms of the bare photon and emitter states as
$|\tilde{0} \rangle  = \sum^{\infty}_{k= 0} (c^{\tilde{0}}_{{\rm g},2k} | {\rm g}, 2k \rangle +
	c^{\tilde{0}}_{{\rm e},2k+1} | {\rm e}, 2k+1 \rangle)$.	
Within the RWA this level is flat (gray dashed line).

We consider the system initially prepared in the lowest energy dressed state $|\tilde{0} \rangle$, which is the system eigenstate closest to $| {\rm g},0 \rangle$. The system can be prepared in the desired state  by exciting the atom with a $\pi$ pulse slow as compared to the inverse of $2 \omega_{0}$ \cite{DeLiberato2} as shown in subsequent numerical calculations (Fig.\ 3d). 
Let us discuss the spontaneous decay of the initial state $|{\rm I_ a} \rangle = |\tilde{0} \rangle$, keeping in mind that spontaneous  transitions induced by a reservoir occur among eigenstates of the total Hamiltonian (environment induced superselection of energy eigenstates \cite{ZurekPRL99}).
For zero or small coupling rates $\Omega_{\rm R}$, the initial state $| \tilde{0} \rangle$ reduces to $| {\rm g},0 \rangle$ and standard spontaneous emission of a photon (at energy $\omega_{\rm gs}$) in the external electromagnetic modes, associated with the emitter transition $|{\rm g}\rangle \to \sigma_{\rm sg} |{\rm g}\rangle = |{\rm s}\rangle$,  occurs at a rate $\gamma_{\rm gs}$ (fixed by the dipole moment of the transition).  In USC, the initial state is $|{\rm I_a} \rangle = | \tilde{0} \rangle$
and possible final states are $|{\rm F}_i \rangle =| {\rm s},2i \rangle$, with transition amplitude   $ \propto \langle {\rm F}_i| \sigma_{\rm sg} |{\rm I_a} \rangle = c^{\tilde{0}}_{{\rm g}, 2i}$. 
Of course spontaneous transitions occurs only if the final states have lower energy than the initial one ($\omega_{\rm s} + 2 i \omega_0 < \omega_{\tilde 0}$).
For $i = 0$, the final state contains no cavity photons as in ordinary spontaneous emission.
The contribution with $i=1$ provides the next dominant term. Hence the spontaneous emission of a  photon not in the cavity mode at a rate $\gamma_{\rm g s}$ comes together with a flux of cavity photon pairs at a rate    $\approx \gamma_{\rm g s} |c^{\tilde{0}}_{{\rm g},2}|^2$. Final states with four cavity photons are also possible in principle but with a much lower rate $\approx \gamma_{\rm g s} |c^{\tilde{0}}_{{\rm g},4}|^2$.

For describing a realistic system, all the dissipation channels need to be taken into account. We adopt the master equation approach. Yet, owing to the very high ratio $\Omega_{\rm R}/\omega_{\rm 0}$, the
description offered by the standard quantum optical master equation  breaks down \cite{Blais}.
Following Refs. \cite{Blais}, we write the system operators in the system-bath interaction Hamiltonian in a basis formed by the eigenstates of $H$. We consider a $T=0$ temperature environment.
Yet generalization to $T\neq0$ environments is straightforward. By applying the standard Markov approximation and tracing out the reservoirs degrees of
freedom , we arrive at the master equation \cite{Blais,RidolfoPRL2011},
$    \dot\rho(t) = i [\rho(t), H] + \sum_c \mathcal{L}_{c}\rho(t)$,
where $\mathcal{L}_{c}$ is a Liouvillian superoperator describing the cavity ($c=0$) and the material system losses 
($c = {\rm e} \to {\rm g}$, and  ${\rm g} \to {\rm s}$) \cite{suppl},
(for details look at Supplementary Informations).
According to the input-output relations in the ultrastrong coupling regime \cite{RidolfoPrl2012}, when the frequency of the emitted photons $\omega \approx \omega_0$, the destruction operator for the output field escaping a single port resonator can be espressed as 
$a_{\rm out}(t) = a_{\rm in}(t) - \sqrt{\gamma_0}\, X^+(t)$. The output cavity photon rate which can be detected in photodetection experiments
is given by the mean value $\Phi_{\rm out} =\langle a^\dag_{\rm out} a_{\rm out}\rangle$. If the input is in the vacuum state as in the present case,
$\Phi_{\rm out}(t) = \gamma_0 \langle X^-(t)\; X^+(t) \rangle$. In circuit-QED systems, a particularly well suited technology for
observing the SCE \cite{suppl}, this normal order correlation function can be measured by using quadrature amplitude detectors \cite{Bozyigit}.
The results of a full numerical demonstration including the cavity and the emitter losses are shown in Fig. 3. Figure 3a displays the numerically calculated time evolution of the mean cavity number of physical photons $\langle X^-(t) X^+(t) \rangle = {\rm Tr}[X^- X^+\, \rho(t)]$ for different coupling strengths $\Omega_{\rm R}/\omega_0 = 0.3$ (red line), 0.4 (blue), 0.6 (black). Calculations have been performed at zero detuning  and  by using $\gamma_{\rm eg}= \gamma_0 = \gamma_{\rm gs}  = 2 \times 10^{-2} \omega_0$, $\omega_{\rm gs} = 3.5 \omega_{0}$.
The system is initially prepared in the electrodynamics vacuum state (the lowest energy state of the interacting sector) $| \tilde 0 \rangle$. Then the spontaneous decay of such state produces an output stream of cavity photons: the unlocked virtual photons that  can now be detected. Such photon stream is the signature of the SCE.
The signal rapidly grows and reaches a maximum value before decaying exponentially due to cavity losses. As expected from the previous analysis, the signal increases with increasing $\Omega_{\rm R}/\omega_0$, as a consequence of the  buildup of $c^{\tilde{0}}_{{\rm g},2}$.
Considering a produced maximum photon number $\langle X^- X^+ \rangle \approx 10^{-2}$ (see Fig.\ 3a) and $\gamma_0 = 2 \times 10^{-2} \omega_0$ with $\omega_0 = 10$ GHz, the SCE will give rise to a peak output photon flux $\Phi^{\rm peak}_{\rm out} \approx 1 \times 10^7$ photons per second. 
Such a photon rate corresponds to a quite low emission power $\hbar \omega_0\; \Phi^{\rm peak}_{\rm out}$ which however can be detected with existing technology \cite{Mariantoni,Bozyigit}. The signal can be significantly enhanced by considering resonators coupled to more than one artificial atom (see panel 3c).
We also notice that, for a typical system temperature $T = 20$ mK, the cavity mode energy $\omega_{0} \approx 10$ GHz is much larger than $K T$ (where $K$ is the Boltzmann constant) and as a consequence the  photon flux (originating from the previously virtual photons) is much higher than that arising from the thermal occupation of the resonator. 
In the absence of USC ($\Omega_{\rm R}/\omega_0 \ll 1$) or applying the RWA the resulting photon rate would be negligible or zero respectively. The detection of a photon flux escaping the cavity is the main signature of the SCE. 
Panel 3b shows calculations of $\langle X^- X^+ \rangle$ for different spontaneous emission decay rates
($\gamma_{\rm gs}/ \omega_0 = 10^{-2}$ (green curve),   $1.5 \times 10^{-2}$ (black), $3 \times 10^{-2}$ (blue), $4 \times 10^{-2}$ (red)) obtained by artificially dropping cavity losses ($\gamma_0 = 0$). We also used a coupling rate $\Omega_{\rm R}/\omega_0 = 0.6$ and $\gamma_{\rm eg}= \gamma_{\rm gs}  = 2 \times 10^{-2} \omega_0$.
In the absence of cavity losses, the mean photon number reaches a maximum value which does not depend on $\gamma_{\rm gs}$.
This result puts forward that the phenomenon here investigated is intrinsically different from the DCE where the emitted photon rate depends strongly on the modulation frequency.
\begin{figure}[!ht]  
\includegraphics[height= 58 mm]{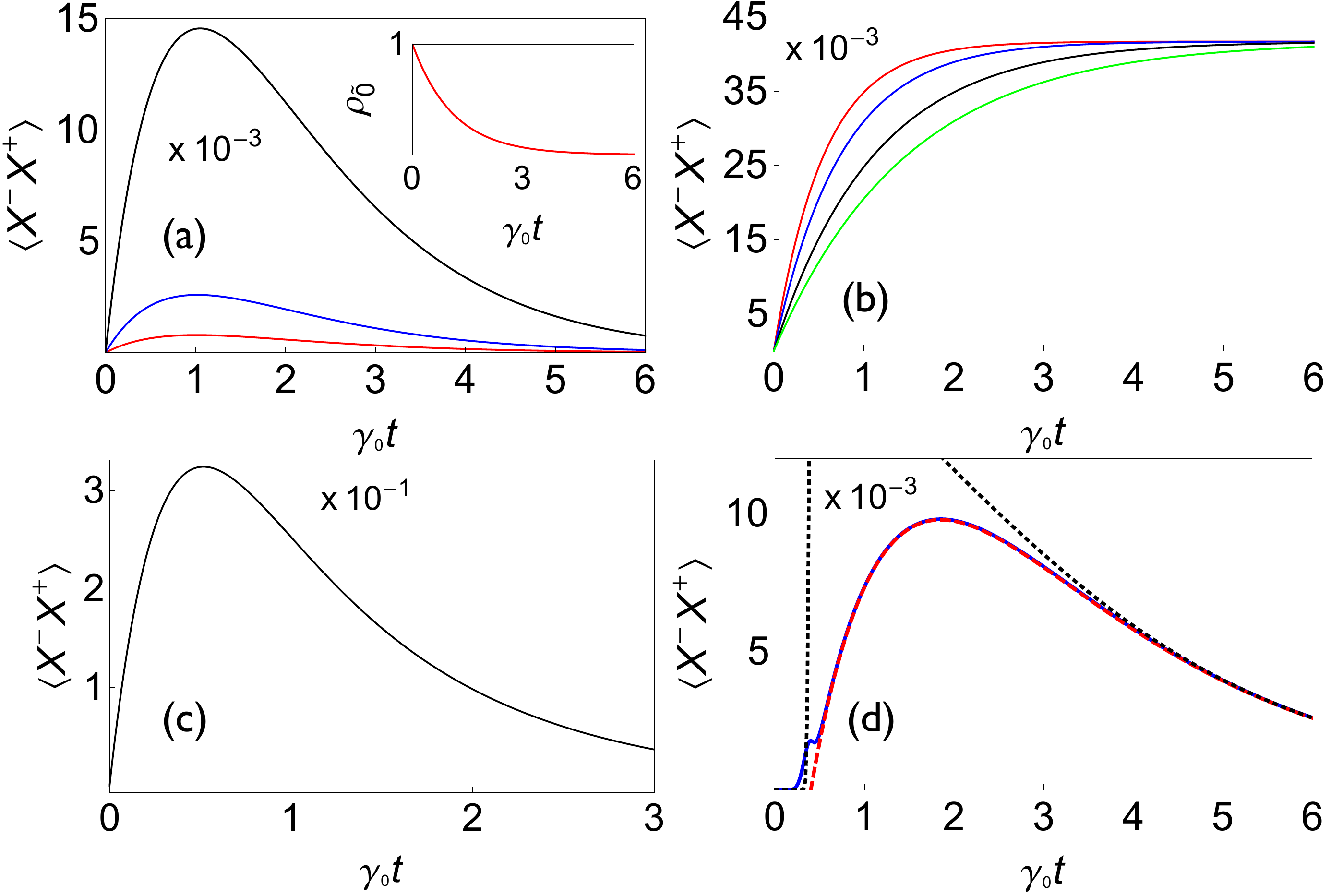}  
\caption{(Color online)
Dynamics of released cavity photons. (a) Time evolution of the intracavity mean photon number  $\langle X^- X^+ \rangle$ calculated for different coupling strengths $\Omega_{\rm R}/\omega_0 = 0.3$ (red line), 0.4 (blue), 0.6 (black). Larger signals corresponds to larger couplings. The inset in panel (a) shows the population dynamics of level $| \tilde{0} \rangle$. (b) $\langle X^- X^+ \rangle$ for different spontaneous emission decay rates $\gamma_{\rm gs}$ (see text) obtained by artificially dropping cavity losses (the signal rises more rapidly for larger $\gamma_{\rm gs}$). (c) Time evolution of $\langle X^- X^+ \rangle$  calculated for two equal artificial atoms with $\Omega_{\rm R}/\omega_0$ = 0.65. (d) Time evolution of $\langle X^- X^+ \rangle$ after  initial state preparation by a $\pi$ pulse with $\sigma = 5/\omega_0$ (continuous blue line), with $\sigma = 1.7/\omega_0$ (dotted black line) and comparison with the dynamics originating from the initial state $| \tilde{0} \rangle$ (dashed red line) for a single three level system.
} 
\label{fig:corrp}
\end{figure}
Figure 3c displays the time evolution of the mean cavity number of physical photons $\langle X^-(t) X^+(t) \rangle$  calculated for two identical artificial atoms.  Calculations have been performed by using $\gamma_{\rm eg}= \gamma_{\rm gs}  = 2 \times 10^{-2} \omega_0$, $\gamma_0 =  1 \times 10^{-2} \omega_0$, $\Omega_{\rm R} = 0.65 \omega_0$ and show that, including additional artificial atoms, significantly enhances the signal. The peak intracavity photon number in Fig.\ 3c largely exceed the sensitivity of circuit QED measurements  ($\approx 0.06$ \cite{Fink}).
Results displayed in Fig.\ 3(a-c) have been obtained preparing the system  in the electrodynamics vacuum state $| \tilde{0} \rangle$ (the lowest energy state of the interacting sector). The simplest way to prepare the system in such a state is to excite the atom with a $\pi$ pulse (with central frequency $\omega = \omega_{\tilde{0}s}$) having a time-width larger than $1/(2 \omega_0)$  and lower than  $1/ \gamma_{\rm g s}$. The excitation is described by the following Hamiltonian term $H_{\rm p}= (\sqrt{\pi/2 \sigma^2} e^{-t^2/(2*\sigma^2)}\cos{\omega t}(\sigma_{gs}+ \sigma_{sg})$. 
Panel 3d compares the dynamics of $\langle X^-(t) X^+(t) \rangle$ after pulse excitation with $\sigma = 5/\omega_0$ (dashed line) with the ideal dynamics obtained starting from the initial state $| \tilde{0} \rangle$. The agreement is very good. On the contrary, if the pulse is too short ($\sigma = 1.7/\omega_0$), the signal (dotted line) strongly differs due to fast modulation induced radiation analogous to the DCE \cite{DeLiberato2}.
Other more sophisticated preparation schemes can be adopted shifting adiabatically the artificial atom levels by an external magnetic flux.

The release of virtual photon pairs, present in  $|\tilde{0} \rangle$, satisfies energy conservation.
The energy of the radiated cavity photons, induced by the spontaneous decay of $|\tilde{0} \rangle$, comes at the expense of the 
energy of the quanta emitted into the reservoir (e.g. the spontaneously emitted photons in the external electromagnetic modes).
In the present case, back-reaction effects can be investigated by calculating the emission spectrum of spontaneously emitted photons directly in the external electromagnetic modes: $S(\omega) = 1/(2 \pi) \int_{-\infty}^\infty dt \int_{-\infty}^\infty dt' \langle \sigma^-(t) \sigma^+(t') \rangle e^{-i \omega (t-t')}$, where $\sigma^{\pm}$ are the frequency positive (negative) components of the polarization operator $\sigma_{gs} + \sigma_{sg}$. The spectrum $S(\omega)$ of spontaneously emitted photons, calculated for $\Omega_{\rm R}/\omega_0 = 0.6$ (other parameters are the same as those used for Fig.\ 3a), is displayed in Fig.\ 4a. In the absence of interaction (dashed line), the spectrum consists of a single Lorentzian peak centered at energy $\omega_{\rm gs}$ corresponding to ordinary spontaneous emission. In USC the main peak is red-shifted at energy $\omega_{{\tilde 0}{\rm s}}$ and  a second peak at lower energy, centered at
$\omega_{{\tilde 0}{\rm s}} - 2 \omega_0$ (see Fig.\ 4a), appears.
This peak shows that pair creation is associated to the spontaneous emission of an outside photon of lower energy $\omega_{\tilde 0s} - 2 \omega_0$. 
The observation of the lower peak in the spontaneous emission spectrum of the emitter would be an additional signature of the SCE.
The higher peak at energy $\omega_{\tilde 0s}$ originates from events where pair creation is absent.
Conspicuous information on the ongoing physics can be obtained studying the statistics of the emitted photons. 
\begin{figure}[!ht]  
\includegraphics[height= 58 mm]{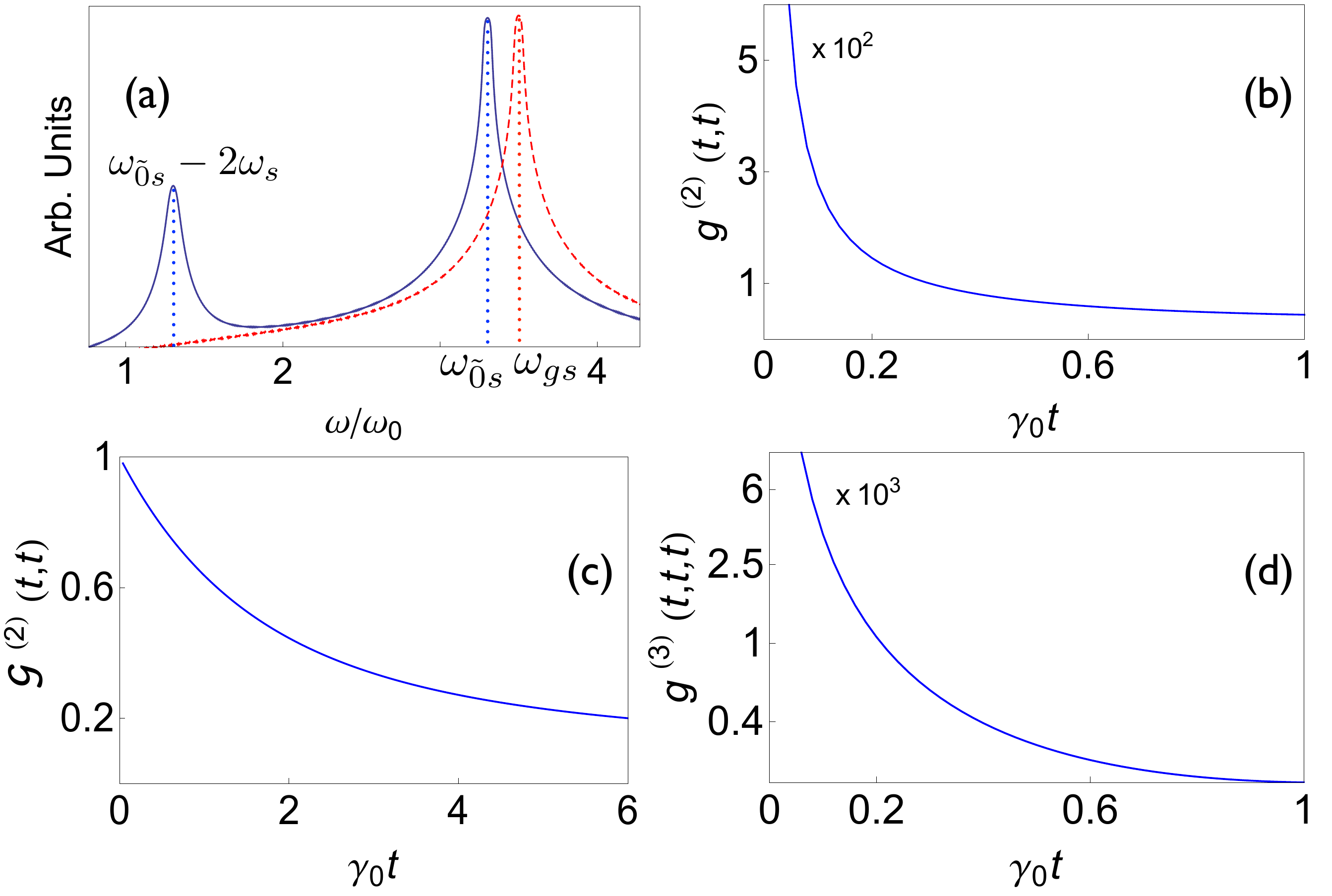}  
\caption{(color online)
(a) Spectrum $S(\omega)$ of spontaneously emitted photons into not-confined electromagnetic modes.
(b) Normalized second order correlation function $g^{(2)}(t)$. (c) ${\cal G}^{(2)}(t,t) = \langle X^-(t) X^+(t) \rangle\, g^{(2)}(t,t)$. (d) Normalized third-order correlation function $g^{(3)}(t)$. } 
\label{fig:4}
\end{figure}
The equal-time normalized second order correlation function for cavity photons, $g^{(2)}(t) = \langle X^-(t) X^-(t) X^+(t)X^+(t) \rangle/\langle X^-(t) X^+(t) \rangle^2$,
shown in Fig. 4b certifies  a highly super-Poissonian statistics, evidencing that cavity photons are released in pairs.
Such pair correlation can be further confirmed by calculating ${\cal G}^{(2)}(t,t) = g^{(2)}(t)\, \langle X^-(t) X^+(t) \rangle$, shown in Fig.\ 4c. It compares the coincidence rate with the ordinary photodetection rate. When photons are emitted in pairs ${\cal G}^{(2)}(t,t) \approx 1$. The decay of these correlation functions is due to cavity losses which tend to destroy correlations. Finally we calculated the equal-time third order normalized correlation function, related to the probability of detecting two cavity photons and a spontaneous emitted photon in the external light modes,
\begin{equation}
	g^{(3)}(t) = \frac{\langle \sigma^-(t) X^-(t) X^-(t) X^+(t)X^+(t) \sigma^+(t) \rangle}
	{\langle \sigma^-(t) \sigma^+(t) \rangle\langle X^-(t) X^+(t) \rangle^2}\, .
\end{equation}
The obtained huge value of  $g^{(3)}(t,t)$ (see Fig.\ 4d) confirms the emission mechanism outlined above.  
Panels 4b-d  have been obtained by using $\Omega_{\rm R}/\omega_0 = 0.6$ and the other parameters as in Fig.\ 3a.
As far as the circuit-QED experimental implementation is concerned, 
the rate of  microwave photons escaping the resonator as well as their correlations can be measured by using quadrature amplitude detectors \cite{Menzel,Bozyigit}. Moreover  tomography of output light released form a circuit cavity has been demonstrated experimentally \cite{Eichler}.

The effect here described paves the way to direct investigation of the most interesting feature of USC, namely the cavity quantum electrodynamic vacuum state. Generalizations of
our study to  arrays of coupled cavities would form interesting perspectives for future research on virtual photons and pair creation in extended systems \cite{Hartmann, Leib10}. 

\vspace{0.5 cm}
\noindent
{\bf Acknowledgement}
\newline
We are grateful to I. Carusotto for useful suggestions.
MJH acknowledges support by the Emmy Noether Program (DFG) and the CRC 631 (DFG) and useful discussions with Martin Leib.

\end{document}